\documentstyle[aps,epsf]{revtex}

\newcommand{\vpsl}{\vec{p}\!\!\!/}

\preprint{BUHEP-99-14,UCTP-109-99}

\begin{document}
\draft

\title{Schwinger-Dyson approach to color superconductivity
in dense QCD}

\author{Deog Ki Hong}
\address{Department of Physics, Pusan National University,
Pusan 609-735, Korea\thanks{Permanent address.}\\
Physics Department, Boston University,
Boston, MA 02215, USA}

\author{V.A.~Miransky}
\address{Bogolyubov Institute for Theoretical Physics,
252143, Kiev, Ukraine\\
Department of Applied Mathematics,
University of Western Ontario,
London, Ontario N6A 5B9, Canada}

\author{I.A.~Shovkovy\thanks{On leave of absence from
Bogolyubov Institute for Theoretical Physics,
252143, Kiev, Ukraine.}
           and
L.C.R.~Wijewardhana}
\address{Physics Department, University of Cincinnati,
Cincinnati, Ohio 45221-0011}

\date{September 17, 1999}
\maketitle

\begin{abstract}
The problem of color superconductivity in dense QCD is
reconsidered in the improved rainbow approximation to the
Schwinger-Dyson equation. The effect of the unscreened
magnetic modes of gluons on the value of the color
condensate is studied. In particular, it is shown that, at
sufficiently large values of the chemical potential, these
modes lead to the enhancement of the superconducting order
parameter. The interplay between the instanton induced
interaction and the one-gluon induced one in color
superconductivity is discussed.
\end{abstract}


\section{Introduction}
\label{Intro}

Quantum chromodynamics (QCD) has been remarkably successful
in describing the interactions of quarks and gluons at short
distance scales. In this distance region, QCD is a perturbative
theory due to asymptotic freedom \cite{AssFree}. At large
distances, on the other hand, the effective running coupling of
QCD becomes strong and non-perturbative approaches, like sum
rule methods \cite{SumR}, lattice computer simulations
\cite{Latt}, instanton computations \cite{Inst}, and
Schwinger-Dyson equations for Green's functions \cite{SD}
should be utilized to analyze the theory.

There exist, however, some extreme conditions where QCD is
yet to be tested. An example of such an extreme condition is
cold quark matter at high density. There have been attempts
to investigate dense QCD using perturbation theory \cite{Bail},
instanton calculations \cite{ARW1,RSSV1,RSSV2} and lattice
simulations \cite{Hands}. Such dense matter may exist in the
interior of neutron or the so-called strange stars \cite{Wit},
with baryon number densities exceeding a few times the normal
nuclear density $n_0\simeq 0.17~\mbox{fm}^{-3}$. Besides that,
the dense enough quark matter could be created in accelerators
by heavy ion collisions. Therefore, the study of such a system
is not of pure academic interest.

At high density, it is believed that the quarks form a Fermi
surface in a very similar way as the electrons do it in metals.
Following this similarity further, it is natural to ask whether
there is an analogue of the superconductivity in the cold
quark-gluon plasma. As in metals, the presence of the Fermi
surface should considerably increase the density of states
of low-energy quasiparticles. As a result, an arbitrarily
small attractive interaction in the diquark channel would be
sufficient to create the Cooper pairs and, thus, to produce the
(color) superconductivity. In Ref.~\cite{Bail}, it was shown
that an attractive interaction in the color-antitriplet
diquark channel indeed appears. As a result, a non-trivial
superconducting order parameter develops and the color gauge
symmetry $SU(3)_{c}$ gets broken (by the Higgs mechanism) down
to its $SU(2)$ subgroup (in the case of two light quark
flavors).

Recently, interest in the study of the color superconducting
phase considerably increased
\cite{ARW1,RSSV1,RSSV2,ARW2,BR,ABR,SW,CD,LR,EHS,PR,Son,Hong,AKS}.
The renewed interest was triggered by the observation that the
necessary attractive interaction could be exclusively due to
the  instanton effects \cite{ARW1,RSSV1,RSSV2}. In addition,
some other fascinating features of the broken phase, such as
color-flavor locking and a new type of chiral symmetry
breaking \cite{ARW2}, were revealed. From a technical point of
view, it was also demonstrated that the renormalization group
method is an alternative and a very efficient tool in studying
the general properties of dense QCD \cite{EHS,PR,Son,Hong}.

While it was believed that the color superconductivity was
quite well understood at least qualitatively, the recent
results of Refs.~\cite{PR,Son} seem to indicate that there is
an inconsistency in all the previous approaches, based on the
straightforward use of the simplest Nambu-Jona-Lasinio model
and the Bardeen-Cooper-Schrieffer (BCS) type of analysis. The
key observation in Refs.~\cite{PR,Son} is that the long range
interaction, mediated by the unscreened gluon modes of the
magnetic type, may considerably enhance the value of the
superconducting order parameter. Such an enhancement would be
of great importance in studies of neutron star properties. 
Note that a somewhat similar conclusion about enhancement was
also reached in Ref.~\cite{Hong} where the effective low-energy
action of the quark quasiparticles around the Fermi surface
was derived.

Being motivated by that observation, in
this paper, we reanalyze the problem by using the
conventional Schwinger-Dyson (SD) approach. More precisely,
we study the  SD equations in dense QCD in the so-called
improved rainbow approximation in which the full vertices
coincide with the bare ones. Such an approximation includes
the one-loop polarization effects in the gluon propagator
and, thus, is  the simplest non-trivial approximation that
takes into account the screening effects. Our expression
for the fermion gap derived in the SD approach turns out to be essentially the
same as in Ref.~\cite{Son} where the renormalization group
method was mainly used.

This paper is organized as follows. In Sec.~\ref{Model}, we
present the general structure of the quark propagator that
accommodates the color superconductivity. In
Sec.~\ref{VacPol}, we discuss the structure of the gluon
propagator with the one-loop  polarization effects taken into
account. The role of the Meissner effect in the color
superconducting phase of QCD is discussed in
Sec.~\ref{meissner}. Then, in Sec.~\ref{SD-Eq}, we derive the
SD equation (in the improved rainbow approximation) for the
quark propagator and present an analytical estimate for the
solution. The discussion of the results is given in the
concluding Section~\ref{Concl}. In Appendix~\ref{appA}, some
formulae used in the derivation of the gap equation are
considered. And, in Appendix~\ref{appB}, we present an
approximate analytical solution to the SD equation.

\section{The Model and Notation}
\label{Model}

Here we consider QCD with two light quark flavors ($u$ and
$d$) in the fundamental representation of the  $SU(3)_{c}$
color gauge group. At sufficiently large values of the chemical
potential, the current masses of the quarks  can be
neglected\footnote{As is argued in Ref.~\cite{ARW2}, the  strange
quark is also sufficiently light to be included in the model.
However, we are not interested here in the specific effects of
the third flavor. Our prime goal here is to clarify the role of
the long range gluon interaction and, thus, it is sufficient
for our purposes to deal with the simplest model of two light
flavors.}. As a consequence, the model is invariant under the
(global) chiral $SU(2)_L \times SU(2)_R$ transformations.

Following the findings of Ref.~\cite{Bail}, we are interested
in studying the diquark condensates $\varepsilon^{ij}
\varepsilon_{ab3} \langle (\psi_{a}^{i})^{T} C \psi_{b}^{j}
\rangle $ and $\varepsilon^{ij} \varepsilon_{ab3} \langle
(\psi_{a}^{i})^{T} C \gamma_{5}\psi_{b}^{j}\rangle$,
responsible for the color superconductivity (here $a$, $b$ and
$i$, $j$ are the color and flavor indices, respectively). We
note, that in perturbation theory the two condensates are
related by $U_{A}(1)$ symmetry. Nonperturbatively, however,
the instanton effects are expected to break the latter
symmetry \cite{ARW1,RSSV1}.

In the presence of a large chemical potential, it is
commonly assumed that the quarks near the Fermi surface (or
rather low-energy quasiparticles) are weakly coupled due to
asymptotic freedom \cite{ColPer}. Then, it is natural to
analyze the dynamics of diquark pairing by using the SD
equations with a perturbative kernel\footnote{Note that
according to Ref.~\cite{EHS}, the effect of instantons could
be negligible compared to the one-gluon exchange at
sufficiently large chemical  potential.} \cite{Hong}. At
this point, it is appropriate to mention that the realistic
value of the chemical potential for the quark matter in the
interior of a neutron star is expected to be somewhere in
between 200 and 700 MeV (so that the baryon number density
is between about $n_0$ and $10n_0$). Then, the
corresponding value of the coupling constant $\alpha_{s}$
should be between about 0.4 and 0.8. At these values of
$\alpha_{s}$, our calculations, based on the perturbative
kernel, may just start to fail. At the same time, since the
values of the coupling constant are not too large, it is
plausible that the orders of the magnitudes of the main
results could still be trusted. We will return to the
discussion of this point in more detail in Sec.~\ref{Concl}.

Here we should mention that the SD equation approach to the
study of color superconductivity in dense QCD was also used
in Refs.~\cite{Bail,LR,Hong,II}. In this paper, however, we
pay special attention to the effect of the unscreened
magnetic modes of gluons.

Instead of working with the standard four component Dirac
spinors, in our analysis below, it is convenient to introduce
the following eight component Majorana spinors:
\begin{equation}
\Psi=\frac{1}{\sqrt{2}}\left(\begin{array}{c}\psi \\
       \psi^{C}
   \end{array}\right), \quad
\psi^{C}=C\bar{\psi}^{T},
\label{Psi}
\end{equation}
where $C$ is a charge conjugation matrix, defined by
$C^{-1} \gamma_{\mu}C =-\gamma_{\mu}^{T}$ and $C=-C^{T}$.
In the new  notation, the inverse fermion propagator (defined
over the  true vacuum) that accommodates the possibility of
the diquark  condensation reads
\begin{equation}
\left(G(p)\right)^{-1}=-i\left(\begin{array}{cc}
A(p)[(p_0+\mu)\gamma^{0}+B(p)\vpsl ]& \Delta\\
\tilde{\Delta} & A(p)[(p_0-\mu)\gamma^{0}+B(p)
\vpsl ] \end{array}\right),
\label{G-inv}
\end{equation}
where $\tilde{\Delta}=\gamma^0\Delta^{\dagger}\gamma^0$ and
$\Delta_{ab}^{ij}\equiv\varepsilon^{ij}\varepsilon_{ab3}
\Delta_{3}$ (note that $\Delta_{3}$ is a matrix in the
Dirac space). By definition, $\vpsl = -\vec{p} \cdot
\vec{\gamma}$.

Now, after inverting the expression in Eq.~(\ref{G-inv}),
we arrive at the following propagator:
\begin{eqnarray}
G(p)&=&i \left(\begin{array}{cc}
R_{1}(p)^{-1} &
-\left[(p_0+\mu)\gamma^{0}+B(p)\vpsl \right]^{-1}
A(p)^{-1} \Delta R_{2}(p)^{-1}\\
-\left[(p_0-\mu)\gamma^{0}+B(p)\vpsl \right]^{-1}
A(p)^{-1} \tilde{\Delta} R_{1}(p)^{-1} &
R_{2}(p)^{-1} \end{array}\right),
\label{G}
\end{eqnarray}
where
\begin{mathletters}
\begin{eqnarray}
R_{1}(p)&=& A(p)
\left[(p_0+\mu)\gamma^{0}+B(p)\vpsl \right]
-\Delta \left[(p_0-\mu)\gamma^{0}
               +B(p)\vpsl \right]^{-1}
A(p)^{-1} \tilde{\Delta},\\
R_{2}(p)&=&A(p) \left[(p_0-\mu)\gamma^{0}
               +B(p)\vpsl \right]
-\tilde{\Delta}
\left[(p_0+\mu)\gamma^{0}+B(p)\vpsl \right]^{-1}
A(p)^{-1}\Delta .
\end{eqnarray}
\end{mathletters}
In Sec.~\ref{SD-Eq}, we use the expression for the propagator
in Eq.~(\ref{G}) as an anzatz for the solution to the SD
equation. Notice that, with the given choice of the
condensate, the most general structure of the wave function
renormalizations $A(p)$ and $B(p)$ (in flavor and color
spaces) has to be as follows: $A_{ab}^{ij}(p) =a(p)
(\delta_{ab} -\delta_{a3} \delta_{b3}) \delta^{ij} +\bar{a}(p)
\delta_{a3} \delta_{b3} \delta^{ij}$ and  $B_{ab}^{ij}(p) =b(p)
(\delta_{ab} -\delta_{a3} \delta_{b3}) \delta^{ij} +\bar{b}(p)
\delta_{a3} \delta_{b3} \delta^{ij}$, where $a(p)$,
$\bar{a}(p)$ and  $b(p)$, $\bar{b}(p)$ are some scalar
functions.

\section{Vacuum Polarization in the Gluon Propagator}
\label{VacPol}

In this section we discuss the vacuum polarization effects in
the gluon propagator.

The calculation of the polarization tensor was performed by
others \cite{Vija,Heinz,Manuel} and we are not going to repeat
it here. Instead we present the final result and comment on its
essential features.

Let us start from the expression for the inverse gluon
propagator in a covariant gauge with polarization effects
taken into account. Its explicit form is given by
\begin{equation}
\left({\cal D}_{\mu\nu}^{AB}(k_0,\vec{k})\right)^{-1}=
i\delta^{AB} k^2 P^{\perp}_{\mu\nu}
+i\delta^{AB} \frac{k^2}{ d } P^{\parallel}_{\mu\nu}
+i\delta^{AB} \Pi_{\mu\nu}(k_0,\vec{k}).\label{inv-Da}
\end{equation}
where the color indices $A, B=1,2,\dots,8$, and
the standard projection
operators $P^{\perp}_{\mu\nu}$ and $P^{\parallel}_{\mu\nu}$,
\begin{equation}
P^{\perp}_{\mu\nu}= g_{\mu\nu}-\frac{k_{\mu}k_{\nu}}{k^2},
\quad \mbox{and} \quad
P^{\parallel}_{\mu\nu}=\frac{k_{\mu}k_{\nu}}{k^2},
\end{equation}
were introduced. The gauge fixing parameter $d$
in Eq.~(\ref{inv-Da}) is arbitrary.

At a finite density of quarks, the Lorentz symmetry is
explicitly broken and, as a result, the polarization tensor
$\Pi_{\mu\nu}$ could also contain the third tensor structure
\cite{Kal}:
\begin{eqnarray}
P^{u}_{\mu\nu}&=& \frac{k_{\mu}k_{\nu}}{k^2}
-\frac{k_{\mu}u_{\nu}+u_{\mu}k_{\nu}}{(u\cdot k)}
+\frac{u_{\mu}u_{\nu}}{(u\cdot k)^2}k^2,
\end{eqnarray}
where $u_{\mu}=(1,0,0,0)$. This last tensor, similarly to
$P^{\perp}_{\mu\nu}$, is transverse,
$k^{\mu}P^{u}_{\mu\nu}=0$. The set of all the three
tensors, introduced so far, satisfy the following set of
multiplication rules:
\begin{mathletters}
\begin{eqnarray}
P^{\perp}P^{\perp}=P^{\perp}, \quad &
P^{\parallel}P^{\parallel}=P^{\parallel}, &\quad
P^{u}P^{u}=\left(\frac{k^2}{(u\cdot k)^2}-1\right)P^{u}, \\
P^{\perp}P^{\parallel}=P^{\parallel}P^{\perp}=0, \quad &
P^{\parallel}P^{u}=P^{u}P^{\parallel}=0, &\quad
P^{u}P^{\perp}=P^{\perp}P^{u}=P^{u},
\end{eqnarray}
\end{mathletters}
where we assume that the operator products are defined by the
appropriate contractions involving the Minkowski metric. While
working with the gluon propagator, it is much more convenient
to use the following set of mutually
orthogonal projection operators:
\begin{mathletters}
\begin{eqnarray}
O^{(1)}=P^{\perp}+\frac{(u\cdot k)^2}{(u\cdot k)^2-k^2} P^{u},
&\quad &
O^{(2)}=-\frac{(u\cdot k)^2}{(u\cdot k)^2-k^2} P^{u}, \quad
O^{(3)}= P^{\parallel},\\
O^{(1)}O^{(1)}=O^{(1)}, &\quad &
O^{(2)}O^{(2)}=O^{(2)}, \quad
O^{(3)}O^{(3)}=O^{(3)}, \\
O^{(1)}_{\mu\nu}+O^{(2)}_{\mu\nu}+O^{(3)}_{\mu\nu}=g_{\mu\nu},
&\quad & O^{(i)}O^{(j)}=0, \quad \mbox{for} \quad i\neq j.
\end{eqnarray}
\label{eq9}
\end{mathletters}

At finite chemical potential, in the one-loop approximation,
the calculation of the polarization tensor reduces to
evaluating the contribution of three diagrams: with gluons,
quarks and ghosts running in the loop \cite{Vija},
respectively. In the hard dense loop approximation,
the result for the polarization tensor is given by
\cite{Vija,Heinz,Manuel},
\begin{mathletters}
\begin{eqnarray}
\Pi^{00}(k_0, \vec{k}) & = & \Pi_{l}(k_0, \vec{k}) ,\\
\Pi^{0i}(k_0, \vec{k}) & = &
k_0 \frac{k^i}{|\vec{k}|^2} \Pi_{l} (k_0, \vec{k}) , \\
\Pi^{ij}(k_0, \vec{k}) & = & \left
( \delta^{ij}- \frac{k^i k^j}{|\vec{k}|^2} \right)
\Pi_{t} (k_0,\vec{k})+ \frac{k^i k^j} {|\vec{k}|^2}
\frac{k_0^2}{|\vec{k}|^2} \Pi_{l} (k_0, \vec{k}),
\end{eqnarray}
\label{Pi}
\end{mathletters}
where
\begin{eqnarray}
\Pi_{l}(k_0,\vec{k})&=&2M^2\left(\frac{k_0}{2|\vec{k}|}
\ln\left|\frac{k_0+|\vec{k}|}{k_0-|\vec{k}|}\right|-1
-i\pi\frac{k_0}{2|\vec{k}|}\theta(-k^2) \right),\\
\Pi_{t}(k_0,\vec{k})&=&M^2-\frac{k^2}{2|\vec{k}|^2}
\Pi_{l}(k_0,\vec{k}).
\end{eqnarray}
Here we use the notation
$M^2=(g_{s}\mu\sqrt{N_f}/2\pi)^2$. What is remarkable about
this result is that it coincides with the polarization
tensor derived in the framework of classical transport
theory of dense Yang-Mills plasma \cite{Heinz,Manuel}.

Notice that the polarization tensor in Eq.~(\ref{Pi}) has
a nonzero imaginary part for space-like gluon momenta.
This imaginary part is responsible for the so-called
Landau damping of the gluon field with space-like momenta.
Also, it is responsible for the quark damping around
the Fermi surface \cite{Vija,VO}. In our analysis below,
however, we neglect the effects of quark damping. Such an
approximation is partly justified by the fact that the
damping rate goes to zero linearly as one approaches the
Fermi surface \cite{VO}.

The gluon polarization tensor in Eq.~(\ref{Pi}) is 
transverse
\begin{equation}
k^{\mu} \Pi_{\mu\nu}(k_0, \vec{k})=0.
\end{equation}
Because of this last property, it is natural to try to 
express the real part of the polarization tensor in
terms of the two transverse  projection operators, $O^{(1)}$
and $O^{(2)}$, introduced earlier. This  turns out to be an
easy task and the result takes the following  nice form:
\begin{equation}
\Pi_{\mu\nu}=-O^{(1)}_{\mu\nu} \Pi_{t}
+2 O^{(2)}_{\mu\nu} \left(\Pi_{t}-M^2\right).
\end{equation}
With this expression at hand, we rewrite the inverse gluon
propagator in Eq.~(\ref{inv-Da}) as
\begin{equation}
\left({\cal D}_{\mu\nu}^{AB}(k_0,\vec{k})\right)^{-1}
=i\delta^{AB} \left(k^2-\Pi_{t}\right) O^{(1)}_{\mu\nu}
+i\delta^{AB} \left(k^2+2\Pi_{t}-2M^2\right) O^{(2)}_{\mu\nu}
+i\delta^{AB} \frac{k^2}{ d } O^{(3)}_{\mu\nu}.
\label{inv-D}
\end{equation}
Then, by making use of the properties of the projection
operators, we could easily invert this expression and
arrive at the final form of the gluon propagator that
includes the (one-loop) screening effects
\begin{equation}
{\cal D}_{\mu\nu}^{AB}(k_0,\vec{k})=-i\delta^{AB}
\frac{1}{k^2-\Pi_{t}} O^{(1)}_{\mu\nu}
-i\delta^{AB} \frac{1}{k^2+2\Pi_{t}-2M^2} O^{(2)}_{\mu\nu}
-i\delta^{AB} \frac{ d }{k^2} O^{(3)}_{\mu\nu}.
\label{D}
\end{equation}
This representation of the gluon propagator, in addition to
its  convenience for calculations, also has another
advantage. It allows  to separate different gluon modes in
a very simple way. Indeed, to make such a separation, we
just need to project the general gluon  state by using one of
the three operators, $O^{(1)}$, $O^{(2)}$ or $O^{(3)}$,
respectively. As a result we come to the following types of
modes: $a^{(m)}_{\mu}=O^{(1)}_{\mu\nu}A^{\nu}$ (magnetic),
$a^{(e)}_{\mu}=O^{(2)}_{\mu\nu}A^{\nu}$ (electric) and
$a^{(\parallel)}_{\mu}=O^{(3)}_{\mu\nu}A^{\nu}$
(longitudinal). The name ``magnetic" here reflects the
point that the  corresponding projection operator
does not have the electric components
$O^{(1)}_{\nu0}=O^{(1)}_{0\nu}=0$. Apparently, the rest of the
transverse modes are the ``electric" modes. As for the
unphysical longitudinal mode, its definition is standard.

In the rest of this Section, we will consider the part of
the gluon propagator connected with the unbroken, $SU(2)$,
subgroup of the $SU(3)_c$, {\em i.e.}, $A, B = 1,2,3$. The
part of the gluon propagator connected with the broken
generators,  $A,B=4,\dots,8$, will be considered in the next
Section.

Because of a specific dynamics of diquark pairing in the
vicinity  of the Fermi surface, it is believed that the
most relevant gluons  mediating the interaction are those
with the space like momenta.  Indeed, when the quarks
around the Fermi surface scatter, their  energy does not
change much, while their momentum could change for  as much
as $2p_F$. Thus, we could mimic the screening effects by
the  following asymptotic form of $\Pi_{t}$ in the region
$|\vec{k}|\gg k_0$:
\begin{equation}
\Pi_{t}(k_0,\vec{k})\simeq 2M^2\frac{k_0^2}
{|\vec{k}|^2}\left(1-\frac{1}{3}\frac{k_0^2}{|\vec{k}|^2}
+\dots\right) +i\pi M^2\frac{k_0}{2|\vec{k}|}
\left(1-\frac{k_0^2}{|\vec{k}|^2}\right)
\theta(|\vec{k}|^2-k_0^2).
\label{asym-2}
\end{equation}
In fact, this is a very good approximation even for the values
of the ratio $k_0/|\vec{k}|$ as large as $0.8$ at which the
deviation of the asymptote from the exact expression is of
order  of $5\%$. If we keep only the leading term in
Eq.~(\ref{asym-2}),  the same magnitude of deviation is
reached at about $k_0/|\vec{k}|  \simeq 0.4$.

By substituting this asymptote into the gluon propagator, we
see that the magnetic mode $a^{(m)}_{\mu}$ produces the
long-range interaction, while the electric mode $a^{(e)}_{\mu}$
gets screened out. In what follows, it is convenient to use the
representation of the gluon propagator in terms of spectral
densities:
\begin{equation}
i{\cal D}_{\mu\nu}^{AB}(k_0,\vec{k})\simeq \delta^{AB}
\lim_{\varepsilon\to 0}\frac{1}{2\pi}\int_{-\infty}^{\infty}
\frac{dz}{k_0-z+i\varepsilon}
\left[\rho_{m}(z,\vec{k})O^{(1)}_{\mu\nu}
+\rho_{e}(z,\vec{k})O^{(2)}_{\mu\nu}
+\rho_{\parallel}(z,\vec{k})O^{(3)}_{\mu\nu}\right] ,
\label{D-spec}
\end{equation}
where
\begin{eqnarray}
\rho_{m}(z,\vec{k})
&\simeq & \frac{\pi M^2 z |\vec{k}|}{|\vec{k}|^6
+(\pi/2)^2 M^4 z^2}\theta(|\vec{k}|^2-z^2), \\
\rho_{e}(z,\vec{k}) &\simeq & 2\pi
\mbox{~sgn}(z)\delta(z^2-|\vec{k}|^2-2M^2),\\
\rho_{\parallel}(z,\vec{k}) & = & 2\pi  d
\mbox{~sgn}(z)\delta(z^2-|\vec{k}|^2).
\label{sp-den}
\end{eqnarray}
Therefore, by making use of the representation in
Eq.~(\ref{D-spec}), we arrive at the following expression
for the gluon propagator in Euclidean space ($k_0=i k_4$):
\begin{equation}
i{\cal D}_{\mu\nu}^{AB}(ik_4,\vec{k})\simeq
-\delta^{AB} \frac{|\vec{k}|}{|\vec{k}|^3+\pi M^2 |k_4|/2}
O^{(1)}_{\mu\nu}
-\delta^{AB} \frac{1}{k_4^2+|\vec{k}|^2+2 M^2 }
O^{(2)}_{\mu\nu}
-\delta^{AB} \frac{ d }{k_4^2+|\vec{k}|^2} O^{(3)}_{\mu\nu},
\label{D-long}
\end{equation}
where $A,B = 1,2,3$. This propagator correctly describes the
gluons in the one-loop approximation in the soft momentum
region. In Sec.~\ref{SD-Eq}, we shall use it in the SD
equation for the quark propagator. As we shall see there,
because of the long  range interaction mediated by the first
term in the propagator in Eq.~(\ref{D-long}), the result for
the order parameter is going to be quite different from that
obtained in the theories with local interactions
\cite{ARW2,EHS}. Notice, that according to the arguments of 
Refs.~\cite{PR,Son}, the magnetic modes of the unbroken $SU(2)$ 
subgroup of $SU(3)_{c}$ should develop no screening even after 
taking into account nonperturbative effects.

\section{The Meissner effect }
\label{meissner}

In this section, we study the role of the Meissner effect
in color superconducting phase of dense QCD. Based on pure
symmetry arguments, it is clear that when the color $SU(3)$
symmetry spontaneously breaks down to $SU(2)$, five out of
eight gluons should get masses by the Higgs mechanism.

To estimate the value of the Higgs-like gluon mass, we
need to calculate  the appropriate contribution to the
vacuum polarization tensor that results from the
non-diagonal term in the fermion  propagator in
Eq.~(\ref{G}). To get a rough estimate of  the mass, we
could completely neglect the effects of the  wave function
renormalizations (in the next Section, we will give
a  justification for
that). Then, we get the following order of magnitude
expression:
\begin{equation}
{\cal P}(0)=M_{0}^2 \simeq
\frac{\alpha_{s}}{\pi^2} \int\frac{d^3 \vec{q} d q_{4}
\mu^2|\Delta ^{(-)}|^2}
{\left[(\mu-|\vec{q}|)^2+q_4^2+|\Delta ^{(-)}|^2\right]^2}
\simeq \frac{\alpha_{s}}{\pi}\mu^2 , \label{higgs}
\end{equation}
where the order parameter $\Delta^{(-)}$ 
is defined in Sec. \ref{SD-Eq} [see Eq.~(\ref{A-d})]. 
This seems to suggest that the five gluons obtain relatively
large masses (of order $\alpha_{s}\mu^2$) due to the
Meissner effect. As we show in a moment, however, this
conclusion is not quite right.

Before going into more details, it is instructive to
remind a few facts about the Meissner effect of the
low-temperature superconductivity in ordinary
(non-relativistic) metals \cite{Lif}. As is well known,
there exist two characteristic scales in the theory: the
coherence  length, $\xi$, and the London magnetic
penetration depth,  $\lambda_L$. For our  purposes here, it
is sufficient to recall that while the coherence length,
as $T\to 0$, is directly related to the value of the
superconducting order parameter ($\xi \sim 1/|\Delta|$),
the London penetration depth is independent of $\Delta$
and is given in terms of the  mass and the density of the
electrons ($\lambda_L \sim \sqrt{m/n_{el}}$).

The London penetration depth $\lambda_L$ coincides with the
actual magnetic penetration depth $\lambda$ only in type II
superconductors with $\lambda_L \gg \xi$: $1/{\lambda_L}$
coincides with the value of the running Higgs-like mass of a
plasmon at zero momentum and the region of small momenta
yields the dominant contribution to the actual penetration
depth in type II superconductors. On the other hand, in type
I superconductors, it is the running Higgs-like mass at
momenta $\Delta \ll k \ll \mu$ that yields the dominant
contribution to $\lambda$. In this case $\lambda=\lambda_P$
where $\lambda_P$ is the Pippard penetration
depth\footnote{We would like to thank S.~Hsu for pointing
our attention to the difference between the definitions of
the London and the Pippard penetration depths.}:
$\lambda_P \sim (\lambda_{L}^2\xi)^{1/3}$ \cite{Lif,FetWal}.

Regarding the cold dense quark matter, it can hardly be
a type II superconductor because, as we shall see from our
final estimate for the gap in Eq.~(\ref{A-d}),
$|\Delta^{(-)}|$ is  $|\Delta ^{(-)}|
\ll M_{0}$, implying that $\lambda_{L}\ll\xi$. Therefore,
most likely, it is a type I superconductor. This means that
the actual penetration depth of a magnetic field is given
not by the London expression but by the Pippard one.
Technically the difference comes from the fact that the
zero-momentum expression in Eq.~(\ref{higgs}), related to
the definition of the London penetration depth, is valid
only in a  very small region of momenta, $|\vec{k}|\ll
|\Delta ^{(-)}|$. In the most important intermediate region,
$|\Delta ^{(-)}|\ll |\vec{k}|\ll \mu$, however, the expression
has the following asymptotic behavior \cite{FetWal}:
\begin{equation}
{\cal P}(k) \simeq M_{0}^2
\frac{|\Delta ^{(-)}|}{|\vec{k}|}.  \label{asy-q}
\end{equation}
Since the relevant momenta for the Meissner effect are of
order $\lambda^{-1} \gg |\Delta ^{(-)}|$, from
Eq.~(\ref{asy-q}), we indeed see that the penetration depth
is equal to $\lambda_{P} \sim (M_{0}^2 |\Delta
^{(-)}|)^{-1/3}$ rather than $\lambda_{L} \sim 1/M_{0}$.

Now, what the role does the Meissner effect play in the
perturbative kernel of the SD equation? To answer this
question, let us consider how the corresponding propagators
of magnetic gluons are going to be modified by the
Meissner effect. Apparently, we could still use the
representation for the propagators as in Eq.~(\ref{D-spec}).
The spectral density, though, should be modified
appropriately:
\begin{equation}
\rho^{(M)}_{m}(z,\vec{k})
\simeq \frac{\pi M^2 z |\vec{k}|} {(|\vec{k}|^3
+M_{0}^2 |\Delta ^{(-)}|)^2 +(\pi/2)^2 M^4
z^2}\theta(|\vec{k}|^2-z^2), \label{Mei-sp}
\end{equation}
where the superscript $M$ stays here for the ``Meissner
effect". After simple calculations, we arrive at the following
expression for the corresponding propagator in Euclidean
space ($k_0=i k_4$):
\begin{equation}
i{\cal D}^{(M)AB}_{\mu\nu} (ik_4,\vec{k})\simeq
-\delta^{AB} \frac{|\vec{k}|}{|\vec{k}|^3
+M_{0}^2 |\Delta ^{(-)}| +\pi M^2 |k_4|/2}  O^{(1)}_{\mu\nu}
+\dots, \qquad\mbox{for}\quad A,B=4,\dots,8,
\label{Mei-D}
\end{equation}
where the ellipsis denote the same electric and longitudinal
contributions as in Eq.~(\ref{D-long}). By comparing the
propagators of magnetic modes in Eqs.~(\ref{D-long}) and
(\ref{Mei-D}), we see that the Meissner effect could be
accounted by the formal replacement $k_4 \to k_4 + c |\Delta
^{(-)}|$ with $c=O(1)$ in the magnetic term of the
five propagators in Eq.~(\ref{Mei-D}). Therefore, it is obvious
that the difference between the propagators in
Eqs.~(\ref{D-long}) and (\ref{Mei-D}) could become important
only in the range of momenta $|k_4| \alt |\Delta ^{(-)}|$ and
$|\vec{k}| \alt(M_{0}^2 |\Delta ^{(-)}|)^{1/3}$. In the next
section, we shall see that the mentioned region of gluon momenta
is not large enough to modify the leading asymptote of the
solution to the gap equation.

\section{Schwinger-Dyson Equation in Dense QCD}
\label{SD-Eq}

In this section we derive the SD equation in dense QCD,
ignoring the masses of quarks. All the necessary
constituents of such an equation were given in the previous
three sections. In the improved rainbow approximation, the
equation reads
\begin{eqnarray}
\left(G_{ab}(p)\right)^{-1}&=&
\left(G^{(0)}_{ab}(p)\right)^{-1}
+4\pi\alpha_{s}\int\frac{d^4 q}{(2\pi)^4}
\left(\begin{array}{cc} \gamma^{\mu} & 0   \\
       0  & -\gamma^{\mu}
  \end{array}\right)
\sum_{a^{\prime},b^{\prime}}\sum_{A=1}^{3}
T^{A}_{a^{\prime}a} G_{a^{\prime}b^{\prime}}(q)
T^{A}_{b^{\prime}b}
\left(\begin{array}{cc} \gamma^{\nu} & 0   \\
       0  & -\gamma^{\nu}
  \end{array}\right)
{\cal D}_{\mu\nu}(q-p) \nonumber \\
&&+4\pi\alpha_{s}\int\frac{d^4 q}{(2\pi)^4}
\left(\begin{array}{cc} \gamma^{\mu} & 0   \\
       0  & -\gamma^{\mu}
  \end{array}\right)
\sum_{a^{\prime},b^{\prime}}\sum_{A=4}^{8}
T^{A}_{a^{\prime}a} G_{a^{\prime}b^{\prime}}(q)
T^{A}_{b^{\prime}b}
\left(\begin{array}{cc} \gamma^{\nu} & 0   \\
       0  & -\gamma^{\nu}
  \end{array}\right)
{\cal D}^{(M)}_{\mu\nu}(q-p),
\label{SD}
\end{eqnarray}
where ${\cal D}_{\mu\nu}(k)$ is the propagator of gluons
that correspond to the unbroken $SU(2)$ subgroup [see
Eq.~(\ref{D-long})], while ${\cal D}^{(M)}_{\mu\nu}(k)$ is the
propagator of those five gluons whose magnetic modes are
modified by the Meissner effect [see Eq.~(\ref{Mei-D})].
Notice, that the overall factor $\delta^{AB}$ is omitted
in the definition of the propagators here. Regarding the rest
of notation, $G_{ab}(p)$ is the full fermion propagator, and
$G^{(0)}_{ab}(p)$ is the perturbative one. With our choice of
the order parameter orientation in the color space,
$\Delta_{ab}\sim \varepsilon_{ab3}$, the explicit form of the
generators of the unbroken color subgroup reads
\begin{equation}
T^{A}=\frac{1}{2}\left(\begin{array}{cc} \sigma^{A} &
                \begin{array}{c} 0 \\ 0 \end{array} \\
       \begin{array}{cc} 0 & 0 \end{array}  & 0
  \end{array}\right), \quad \mbox{where} \quad  A=1,2,3,
\end{equation}
and $\sigma^{A}$ are the Pauli matrices. Then, by making use
of the well known identity for the Pauli matrices, we arrive
at  the following result for the summation over $A$:
\begin{equation}
\sum_{A=1}^{3}T^{A}_{a^{\prime}a}T^{A}_{b^{\prime}b}
=\frac{1}{2}\left(
\delta_{a^{\prime}b}-\delta_{a^{\prime}3}\delta_{b3}
\right) \left(
\delta_{ab^{\prime}}-\delta_{a3}\delta_{b^{\prime}3}
\right)-\frac{1}{4} \left(
\delta_{aa^{\prime}}-\delta_{a3}\delta_{a^{\prime}3}
\right) \left(
\delta_{bb^{\prime}}-\delta_{b3}\delta_{b^{\prime}3}
\right).\label{TaTa}
\end{equation}
By making use of the identity for all eight generators
of $SU(3)$:
\begin{equation}
\sum_{A=1}^{8}T^{A}_{a^{\prime}a}T^{A}_{b^{\prime}b}
=\frac{1}{2}\delta_{a^{\prime}b}\delta_{ab^{\prime}}
-\frac{1}{6}\delta_{aa^{\prime}}\delta_{bb^{\prime}},
\label{TT}
\end{equation}
we also obtain the following one:
\begin{equation}
\sum_{A=4}^{8}T^{A}_{a^{\prime}a}T^{A}_{b^{\prime}b}
=\frac{1}{12}\delta_{aa^{\prime}}\delta_{bb^{\prime}}
+\frac{1}{2}\left(
\delta_{a^{\prime}b}\delta_{a3}\delta_{b^{\prime}3}
+\delta_{ab^{\prime}}\delta_{a^{\prime}3}\delta_{b3}
\right)
-\frac{1}{4}\left(
\delta_{aa^{\prime}}\delta_{b3}\delta_{b^{\prime}3}
+\delta_{bb^{\prime}}\delta_{a3}\delta_{a^{\prime}3}
\right) -\frac{1}{4}
\delta_{a3}\delta_{b3}\delta_{a^{\prime}3}
\delta_{b^{\prime}3},
\label{TT-tt}
\end{equation}

Now, upon inserting the identities (\ref{TaTa}) and
(\ref{TT-tt}) along with the expression for the fermion
propagator (\ref{G}) into  Eq.~(\ref{SD}), we arrive at
the set of integral equations:
\begin{eqnarray}
a(p) \Delta^{(\pm)}(p)&=&\frac{\pi\alpha_{s}}{6}
\int\frac{d^4 q}{(2\pi)^4}\frac{1}{a(q)}
\left(9{\cal D}_{\mu\nu}(q-p)
-{\cal D}^{(M)}_{\mu\nu}(q-p)\right) \nonumber\\
&\times&\left[\frac{\Delta ^{(-)}(q) \mbox{~Tr}
\left(\gamma^{\mu}\Lambda^{(+)}_{q}\gamma^{\nu}
\Lambda^{(\pm)}_{p}\right)}{q_0^2-(b|\vec{q}|-\mu)^2
-|\Delta ^{(-)}|^2}
+\frac{\Delta ^{(+)}(q) \mbox{~Tr}
\left(\gamma^{\mu}\Lambda^{(-)}_{q}\gamma^{\nu}
\Lambda^{(\pm)}_{p}\right)}{q_0^2-(b|\vec{q}|+\mu)^2
-|\Delta ^{(+)}|^2}\right],
\label{gap}\\
(a-1)(p_0+\mu) &\mp & (ab-1)|\vec{p}|
=-\frac{\pi\alpha_{s}}{6}\int\frac{d^4 q}{(2\pi)^4}
\frac{1}{a(q)}\left(3{\cal D}_{\mu\nu}(q-p)
+{\cal D}^{(M)}_{\mu\nu}(q-p)\right) \nonumber\\
&\times&\left[\frac{\left(q_0+b|\vec{q}|-\mu\right)
\mbox{~Tr} \left(\gamma^{0}\gamma^{\mu} \Lambda^{(+)}_{q}
\gamma^{0} \gamma^{\nu} \Lambda^{(\pm)}_{p}\right)}
{q_0^2-(b|\vec{q}|-\mu)^2 -|\Delta ^{(-)}|^2}
+\frac{\left(q_0-b|\vec{q}|-\mu\right) \mbox{~Tr}
\left(\gamma^{0}\gamma^{\mu}\Lambda^{(-)}_{q}\gamma^{0}
\gamma^{\nu}\Lambda^{(\pm)}_{p}\right)}
{q_0^2-(b|\vec{q}|+\mu)^2-|\Delta ^{(+)}|^2}\right],
\label{a} \\
(\bar{a}-1)(p_0+\mu) &\mp & (\bar{a}\bar{b}-1)|\vec{p}|
=-\frac{2\pi\alpha_{s}}{3}\int\frac{d^4 q}{(2\pi)^4}
\frac{1}{\bar{a}(q)} {\cal D}^{(M)}_{\mu\nu}(q-p)
\nonumber\\
&\times&\left[ \frac{\mbox{Tr}
\left(\gamma^{0}\gamma^{\mu} \Lambda^{(+)}_{q}
\gamma^{0} \gamma^{\nu} \Lambda^{(\pm)}_{p}\right)}
{q_0-\bar{b}|\vec{q}|+\mu}
+\frac{\mbox{Tr}
\left(\gamma^{0}\gamma^{\mu}\Lambda^{(-)}_{q}\gamma^{0}
\gamma^{\nu}\Lambda^{(\pm)}_{p}\right)}
{q_0+\bar{b}|\vec{q}|+\mu}\right],
\label{bar-a}
\end{eqnarray}
Here, $\mbox{Tr}$ denotes the trace in Dirac indices, and
$\Delta^{(\pm)}$ are complex functions of momentum, defined by
either\footnote{Notice that we factored out the renormalization
of the wave function $a(p)$ in the definition of
$\Delta_{3}(p)$.} $\Delta_{3}(p) =a(p) \left[ \Delta ^{(+)}(p)
\Lambda^{(+)}_{p} +\Delta ^{(-)}(p) \Lambda^{(-)}_{p} \right]$
(parity odd order parameter) or $\Delta_{3}(p) =a(p) \gamma_{5}
\left[ \Delta ^{(+)}(p) \Lambda^{(+)}_{p} +\Delta ^{(-)}(p)
\Lambda^{(-)}_{p} \right]$ (parity even order parameter),  while
\begin{equation}
\Lambda^{(\mp)}_{p}\equiv \frac{1}{2}
\left(1\mp \frac{\vec{\alpha}\cdot \vec{p}}{|\vec{p}|}\right)
\label{Lambdas}
\end{equation}
are the free quark (antiquark) on-shell projectors
\cite{PR,SchW,PR2}\footnote{In our notations,
$\Delta^{(-)}$ is the same as $(\Delta_1)^*$ of
Ref.~\cite{SchW}.}.
 
Let us start our analysis by considering the equations for the
wave function renormalizations (\ref{a}) and (\ref{bar-a})
first. In order to get a rough estimate for the wave function
renormalizations, it is sufficient to substitute $a =b
=\bar{a} =\bar{b} =1$ along with $\Delta^{(\pm)}=0$ into the
right hand side of Eqs.~(\ref{a}) and (\ref{bar-a}). Then, it
is straightforward to show that (i) the wave function
renormalization of the temporal and the spatial parts of the
fermion kinetic terms are not equal, and (ii) the one-loop
corrections to $a(p)$ and $b(p)$ develop logarithmic
divergences at $p_4=0$ and $|\vec{p}|=\mu$. As is clear, these
logarithmic divergences at the Fermi surface are going to be
removed when a non-zero order parameter $\Delta ^{(-)}$ is
reintroduced. Notice that, because of the Meissner effect, no
infrared divergences develop in $\bar{a}(p)$ and $\bar{b}(p)$
either. At the end, we obtain the following estimates:
\begin{equation}
a,  b, \bar{a}, \bar{b} \simeq 1+\mbox{Const}\cdot
\alpha_s (\mu) \ln\frac{\mu}{|\Delta ^{(-)}|}.
\label{w-f-r}
\end{equation}
By taking into account the expected value of the order
parameter [see Eq.~(\ref{A-d}) below], we actually see
that all wave function renormalizations are close to $1$
if the coupling constant $\alpha_s (\mu)$ is weak. Therefore,
we conclude that, in the leading order approximation, it is
justified to neglect the wave function renormalization effects.

The fact that $a\approx b\approx \bar{a}\approx \bar{b}
\approx 1$ considerably simplifies the study of the gap
equation (\ref{gap}). However, one still has a rather
complicated set of two coupled integral equations for
$\Delta ^{(-)}$ and $\Delta ^{(+)}$, respectively. At this point,
it is important to notice that only $\Delta ^{(-)}$ defines
the gap in the quasiparticle spectrum around the Fermi
surface. Also, in the leading order approximation,
it is sufficient to keep only those terms on the right hand
side of Eq.~(\ref{gap}) which become singular at the Fermi
surface as $\Delta ^{(-)}\to 0$. Then, we see that the
equation for $\Delta ^{(-)}(p)$ decouples:
\begin{eqnarray}
\Delta ^{(-)}(p)&=&\frac{\pi\alpha_{s}}{6}
\int\frac{d^4 q}{(2\pi)^4}
\left(9{\cal D}_{\mu\nu}(q-p)
-{\cal D}^{(M)}_{\mu\nu}(q-p)\right)
\frac{\Delta ^{(-)}(q) \mbox{~Tr}
\left(\gamma^{\mu}\Lambda^{(+)}_{q}\gamma^{\nu}
\Lambda^{(-)}_{p}\right)}{q_0^2-(|\vec{q}|-\mu)^2
-|\Delta ^{(-)}|^2}.
\label{gap-min}
\end{eqnarray}
We further simplify the gap equation by assuming that
$\Delta ^{(-)}(p) \equiv \Delta ^{(-)}(p_4,\vec{p}_{F})$ with
$\vec{p}_{F}=(0,0,\mu)$. This approximation with the order
parameter being a function of only $p_4$ is partly justified
by the structure of the magnetic (dominant) part of the
perturbative kernel in the SD equation. Indeed, in the
vicinity of the Fermi surface, the dependence of the kernel is
more sensitive to changes in $p_4$ than to changes in
$|\vec{p}|$. This is due to the fact that 
while the dependence on
$p_4$ comes from the linear term in the denominator of the
propagator of the magnetic modes in Eqs.~(\ref{D-long}) and
(\ref{Mei-D}), the dependence on $|\vec{p}|$ comes from the 
cubic term.

In order to linearize Eq.~(\ref{gap-min}), we  substitute
$\Delta^{(-)}(p) \to \Delta ^{(-)} \equiv
\Delta^{(-)}(p_4)|_{p_4=0}$ in the denominator. Then, after
taking the trace over the Dirac indices and switching to the
Euclidean space ($q_0=iq_4$), the  integral over the angular
coordinates in Eq.~(\ref{gap-min}) can be done exactly (see
Appendix~\ref{appA} for details). After performing the
subsequent approximate integration over the absolute value of
the spatial momentum, we finally arrive at
\begin{equation}
\Delta ^{(-)}(p_4)=\frac{\alpha_{s}}{72\pi}
\int_{-\infty}^{\infty} d q_{4}
\frac{\Delta ^{(-)}(q_4)}
{\sqrt{q_4^2+|\Delta ^{(-)}|^2}}
\left(9\ln\frac{2(2\mu)^3}{\pi M^2 |q_4-p_4|}
-\ln\frac{2(2\mu)^3}{2M_0^2|\Delta ^{(-)}|
+\pi M^2 |q_4-p_4|}+12\ln\frac{(2\mu)^2}{2M^2}+12d
\right),
\label{gap-00}
\end{equation}
As is easy to trace back, the first two terms in this
expression come from the interaction mediated by the magnetic
modes: while the first one is connected with the three gluons
of the unbroken $SU(2)$ subgroup, the second term comes from
the interaction mediated by the five gluons subject to the
Meissner effect. The third term in Eq.~(\ref{gap-00}) is
connected with the electric modes, and the last term comes
from the longitudinal modes. By noticing that the main
contribution to the right hand side comes from the region of
momenta $|\Delta ^{(-)}|\ll |q_4| \ll \mu$, it is obvious that
the Meissner effect is of no importance in this gap equation.
Therefore, by  putting $M_0=0$, we arrive at
\begin{equation}
\Delta ^{(-)}(p_4)\simeq \frac{\alpha_{s}}{9\pi}
\int \frac{d q_4 \Delta ^{(-)}(q_4)}
{\sqrt{q_4^2+|\Delta ^{(-)}|^2}}
\ln\frac{\Lambda}{|q_4-p_4|} ,
\label{gap-q4}
\end{equation}
where $\Lambda\equiv  e^{3d/2} (2\mu)^6/(\sqrt{2}\pi M^5)
=e^{3d/2}(4\pi)^{3/2}\mu/\alpha^{5/2}$.

Notice that, because of the absence of the logarithmic
factor in front of the gauge parameter in Eq.(\ref{gap-00}),
the longitudinal gluon modes become relevant only in the
next-to-leading order. Here, however, we keep their
contribution in order to estimate the size of possible
corrections to the magnitude of the order parameter due to
subleading effects [see the discussion following
Eq.~(\ref{A-d})]. The contribution of the electric modes is also 
subleading. However, because of the large logarithm 
$\ln(\mu^2/M^2)\sim \ln(1/\alpha_s)$, their contribution to
the gap is important at asymptotically high density of
quark matter when the coupling $\alpha_{s}(\mu)$ is weak 
[see the discussion after Eq.~(\ref{A-d})]. 

To get a rough estimate of the order of the gap, we could
just put a sharp ultraviolet cut-off at $q_4=\mu$ and then,
assuming that $\Delta ^{(-)}(q_4) \approx
\Delta ^{(-)}=\mbox{Const}$
in this region of energies, we come to the algebraic
equation:
\begin{equation}
1\simeq\frac{2\alpha_{s}}{9\pi}
\int_{|\Delta ^{(-)}|}^{\mu} \frac{d x }{x}
\ln\frac{2(2\mu)^3}{\pi M^2 x}
\simeq\frac{\alpha_{s}}{9\pi}
\ln^2\frac{\Lambda}{|\Delta ^{(-)}|},
\label{angle}
\end{equation}
leading to the following estimate for the gap:
\begin{equation}
|\Delta ^{(-)}|\simeq
\frac{(4\pi)^{3/2}\mu}{\alpha^{5/2}} e^{3d/2}
\exp\left(-3\sqrt{\frac{\pi}{\alpha_{s}}}\right).
\label{last}
\end{equation}
A more rigorous solution of the integral equation
(\ref{gap-q4}) is presented in Appendix~\ref{appB}. The
latter gives a slightly smaller value of the gap:
\begin{equation}
|\Delta ^{(-)}|\simeq
\frac{(4\pi)^{3/2}\mbox{e}\mu}{\alpha^{5/2}} e^{3d/2}
\exp\left(-
\frac{3 \pi^{3/2}}{2^{3/2}\sqrt{\alpha_{s}}}
\right)
=\frac{2^7 \pi^4 \mbox{e}\mu}{ g_{s}^{5}} e^{3d/2}
\exp\left(-\frac{3\pi^2}{\sqrt{2}g_{s}}\right),
\label{A-d}
\end{equation}
where $\mbox{e}=2.718\dots$. 
Notice that the
exponent in this expression is completely determined
by the magnetic modes. The main role of the electric
modes is in changing the power of $g_s$ in the prefactor:
from $g_{s}^{-2}$ to $g_{s}^{-5}$. At last, the contribution
of the longitudinal modes in the prefactor is formally of
order one and subleading, though it is essentially
different from 1 for large values of the gauge parameter
$|d|$. Obviously, a proper consideration of all subleading
effects (vertex corrections, wave function renormalizations,
the Meissner effect, etc.) would result in the cancellation of
the gauge dependence in the order parameter, though it is
hard to show this explicitly. Our consideration here just
indicates that subleading corrections in the prefactor might
be, though of order one, not small.

The exponent and the power of $g_s$ in the prefactor in
expression (\ref{A-d}) coincide with those in the expression
for the gap obtained in Ref. \cite{Son} by using the
renormalization group method. Also, up to an overall factor of
order one, this expression agrees with those in recent papers
Refs. \cite{SchW,PR2} where  the SD equation in dense QCD has
been studied by using somewhat different approaches.

Regarding the validity of the expression in Eq.~(\ref{A-d}),
one has to bare in mind that most of the approximations
made at the intermediate steps are adequate only for small
values of the coupling constant. Therefore, in the most
interesting case of the values of $\alpha_{s}\sim O(1)$, the
estimate in Eq.~(\ref{A-d}) is not very reliable. Still we
believe that it should give a correct order of magnitude for
$|\Delta ^{(-)}|$.

\section{Conclusion}
\label{Concl}

In this paper, we studied the phenomenon of color
superconductivity in dense QCD with two light flavors in
the framework of the SD equation in the improved rainbow
approximation. Since our approach is based on the SD
equations with a perturbative kernel, it may provide a
reliable description of the system only at sufficiently
large values of the chemical potential (say, larger than
about a few GeV). Remarkably, the latter restriction on the
range of validity is also sufficient for suppressing the
non-perturbative contributions given by the instanton
effects \cite{RSSV2}.

Regarding our analytical dependence of the gap on the
coupling constant presented in Eq.~(\ref{A-d}), a few
comments are in order. First of all, because of
the large prefactor in Eq.~(\ref{A-d}),
the magnitude of the gap for realistic values of the
chemical potential (200 to 700 MeV) and the coupling
constant (0.8 to 0.4, respectively) can be as
large as of order 100 MeV  
(see Fig.~\ref{fig-del}). Therefore, it is of the same
order as the non-perturbative contribution due to the
instantons.  When the quark density is sufficiently large,
however, the instanton effects become negligible and the
order parameter is primarily determined by the gluon
interaction. Indeed, the key point to notice is that the
expression in Eq.~(\ref{A-d}) contains $\alpha_s^{-1/2}$ in
the exponent in contrast to $\alpha_s^{-1}$ as appears in
models with local four-fermion interactions induced, for
example, by screened gluons or by instantons. Therefore, at
small values of $\alpha_s$, the estimate in Eq.~(\ref{A-d})
is much larger compared to the estimates from four-fermion
models.

Thus, our analysis clearly shows that the long-range
interaction, mediated by the unscreened magnetic modes of
gluons, is responsible for the enhancement of the
superconducting gap in the asymptotic region of high
densities. This confirms the general conclusions of
Ref.~\cite{Son} based primarily on renormalization group
methods.

It is natural to conjecture that expression (\ref{A-d}),
derived in the improved rainbow approximation, yields the
exact essential singularity for the gap at $\alpha_s(\mu)$=0
(corresponding to the chemical potential $\mu$ going to
infinity). In other words, the improved rainbow
approximation can be a good leading approximation for
large values of the chemical potential when the instanton
contributions are suppressed. Indeed, the relevant region of
momenta in the gap equation (\ref{gap-00}) is $|\Delta
^{(-)}| \ll |q_4| \ll \mu$. Since $\alpha_s(\mu)\sim 1/\ln
(\mu/ \Lambda_{QCD})$, expression (\ref{A-d}) implies that
the gap $|\Delta ^{(-)}|$ goes to infinity with $\mu$.
Therefore the running coupling is small in that relevant
region. This suggests that the exponent in expression
(\ref{A-d}) for the gap is exact (on the other hand, the
prefactor can be influenced by higher order corrections in
the kernel). This issue deserves further study.

We note that the two approaches, {\em i.e.}, using either
the perturbative interaction or the non-perturbative
instanton mediated one, seem to be complimentary ways of
describing the intermediate region of dense QCD. While the
standard perturbative approach works well at large chemical
potential, the instanton  approach gives a more appropriate
description at small chemical  potential.

\begin{acknowledgments}
We thank R. Pisarski for discussions having clarified some
inconsistencies in the original version of the paper.
I.A.S. would like to thank the hospitality of Aspen Center
for Physics where a part of this work has been done. He also
acknowledges useful discussions with W.~Bardeen, P.~Damgaard,
F.~P.~Esposito, E.~Gorbar and M.~Jarrell. One of us (D.K.H.)
thanks S.~Hsu for useful discussions and wishes to acknowledge
the financial support of the Korea Research Foundation made in
the program year of 1998 (1998-15-D00022). The work of I.A.S.
and L.C.R.W. was supported by the U.S. Department of Energy
Grant  No. DE-FG02-84ER40153.
\end{acknowledgments}

\appendix

\section{Formulae used in derivation of the gap equation}
\label{appA}

In derivation of Eqs.~(\ref{gap}) -- (\ref{bar-a}), we find
it convenient to use the following representations for the
kinetic terms of the quark propagators:
\begin{eqnarray}
(p_0+\mu)\gamma^{0}+B \vpsl &=& \gamma^{0}
\left[(p_0-B |\vec{p}|+\mu) \Lambda^{(+)}_{p}
+(p_0+B |\vec{p}|+\mu) \Lambda^{(-)}_{p} \right],
\label{a-1}\\
(p_0-\mu)\gamma^{0}+B \vpsl &=& \gamma^{0}
\left[(p_0-B |\vec{p}|-\mu) \Lambda^{(+)}_{p}
+(p_0+B |\vec{p}|-\mu) \Lambda^{(-)}_{p} \right],
\label{a-2}\\
\left[(p_0+\mu)\gamma^{0}+B \vpsl \right]^{-1} &=&
\gamma^{0}\left[ \frac{1}{p_0+B |\vec{p}|+\mu} 
\Lambda^{(+)}_{p}
+\frac{1}{p_0-B |\vec{p}|+\mu} \Lambda^{(-)}_{p} \right],
\label{a-3}\\
\left[(p_0-\mu)\gamma^{0}+B \vpsl \right]^{-1} &=&
 \gamma^{0}\left[
\frac{1}{p_0+B |\vec{p}|-\mu} \Lambda^{(+)}_{p}
+\frac{1}{p_0-B |\vec{p}|-\mu} \Lambda^{(-)}_{p} \right],
\label{a-4}
\end{eqnarray}
where the projectors $\Lambda^{(\pm)}_{p}$ are defined in
Eq.~(\ref{Lambdas}) in the main text.

In order to perform the angular integration in the right hand
sides  of the gap equation (\ref{gap}), one first need to
calculate the following two types of traces over the Dirac
indices:
\begin{eqnarray}
\mbox{~tr}\left[\Lambda^{(\pm)}_{p}\gamma^{\mu}
\Lambda^{(\pm)}_{q}\gamma^{\nu}\right] &=&
g^{\mu\nu} (1+t )-2 g^{\mu 0}g^{\nu 0} t
+\frac{\vec{q}^{\mu}\vec{p}^{\nu}+\vec{q}^{\nu}\vec{p}^{\mu}}
{|\vec{q}| |\vec{p}|}+\dots, \label{a-5}\\
\mbox{~tr}\left[\Lambda^{(\pm)}_{p}\gamma^{\mu}
\Lambda^{(\mp)}_{q}\gamma^{\nu}\right] &=&
g^{\mu\nu} (1-t )+2 g^{\mu 0}g^{\nu 0} t
-\frac{\vec{q}^{\mu}\vec{p}^{\nu}+\vec{q}^{\nu}\vec{p}^{\mu}}
{|\vec{q}| |\vec{p}|}+\dots,
\label{a-6}
\end{eqnarray}
where $t =\cos\theta$ is the cosine of the angle between
three-vectors $\vec{q}$ and $\vec{p}$, and irrelevant
antisymmetric terms are denoted by ellipsis.

By contracting these traces with the projectors of the
magnetic, electric and longitudinal types of gluon modes, we
arrive at 
\begin{mathletters}
\begin{eqnarray}
O^{(1)}_{\mu\nu} \mbox{~tr}
\left[\Lambda^{(\pm)}_{p}\gamma^{\mu}
\Lambda^{(\pm)}_{q}\gamma^{\nu}\right] &=& 2 (1+t )
\frac{q^2+p^2-q p(1+t )}
{q^2+p^2-2q pt }, \label{a-7}\\
O^{(1)}_{\mu\nu} \mbox{~tr}
\left[\Lambda^{(\pm)}_{p}\gamma^{\mu}
\Lambda^{(\mp)}_{q}\gamma^{\nu}\right] &=&2 (1-t )
\frac{q^2+p^2+q p(1-t )}
{q^2+p^2-2q pt }, \label{a-8}\\
O^{(2)}_{\mu\nu} \mbox{~tr}
\left[\Lambda^{(\pm)}_{p}\gamma^{\mu}
\Lambda^{(\pm)}_{q}\gamma^{\nu}\right] &=&2 (1-t )
\frac{q^2+p^2+q p(1-t )}{q^2+p^2-2q pt }
- (1-t )\frac{(q+p)^2+(q_4-p_4)^2} {q^2+p^2-2q pt
+(q_4-p_4)^2}, \label{a-9}\\
O^{(2)}_{\mu\nu} \mbox{~tr}
\left[\Lambda^{(\pm)}_{p}\gamma^{\mu}
\Lambda^{(\mp)}_{q}\gamma^{\nu}\right] &=&2 (1+t )
\frac{q^2+p^2-q p(1+t )}{q^2+p^2-2q pt }
-(1+t ) \frac{(q-p)^2+(q_4-p_4)^2}{q^2+p^2-2q pt
+(q_4-p_4)^2}, \label{a-10}\\
O^{(3)}_{\mu\nu} \mbox{~tr}
\left[\Lambda^{(\pm)}_{p}\gamma^{\mu}
\Lambda^{(\pm)}_{q}\gamma^{\nu}\right] &=& (1-t )
\frac{(q+p)^2+(q_4-p_4)^2}
{q^2+p^2-2q pt
+(q_4-p_4)^2}, \label{a-11}\\
O^{(3)}_{\mu\nu} \mbox{~tr}
\left[\Lambda^{(\pm)}_{p}\gamma^{\mu}
\Lambda^{(\mp)}_{q}\gamma^{\nu}\right] &=& (1+t )
\frac{(q-p)^2+(q_4-p_4)^2}
{q^2+p^2-2q pt +(q_4-p_4)^2},
\label{a-12}
\end{eqnarray}
\label{Otr}
\end{mathletters}
where $q\equiv |\vec{q}|$, $p\equiv |\vec{p}|$,  
$q_4\equiv -i q_0$ and $p_4\equiv -i p_0$.

Now, by making use of the explicit expressions (\ref{Otr}), we
easily perform the angular integrations of all types appearing in
the gap equations:
\begin{eqnarray}
&&\int \frac{d\Omega |\vec{q}-\vec{p}|}
{|\vec{q}-\vec{p}|^3+\omega_{l}^3}
O^{(1)}_{\mu\nu} \mbox{~tr}
\left[\Lambda^{(\pm)}_{p}\gamma^{\mu}
\Lambda^{(\pm)}_{q}\gamma^{\nu}\right] = \pi \Bigg[
-\frac{2}{qp}+\frac{(q^2-p^2)^2+\omega_{l}^4}
{\sqrt{3}\omega_{l}^2 q^2 p^2}\arctan
\left(\frac{\sqrt{3}\omega_{l}\mbox{min}(q,p)}
{\omega_{l}^2+|q^2-p^2|-\omega_{l}\mbox{max}(q,p)}\right)
 \nonumber\\
&&+\frac{(q^2-p^2)^2-\omega_{l}^4} {3\omega_{l}^2 q^2 p^2}
\ln\frac{\omega_{l}+|q+p|}{\omega_{l}+|q-p|}
-\frac{(q^2-p^2)^2-\omega_{l}^4} {6\omega_{l}^2 q^2 p^2}
\ln\frac{\omega_{l}^2+|q+p|^2-\omega_{l}|q+p|}
{\omega_{l}^2+|q-p|^2-\omega_{l}|q-p|}+\frac{4}{3qp}
\ln\frac{\omega_{l}^3+|q+p|^3}{\omega_{l}^3+|q-p|^3}
\Bigg],
\label{a-13}
\end{eqnarray}
where $\omega_{l}^3=(\pi/2) M^2 \omega$ and 
$\omega =|q_4-p_4|$. Similarly,
\begin{eqnarray}
&&\int \frac{d\Omega |\vec{q}-\vec{p}|}
{|\vec{q}-\vec{p}|^3+\omega_{l}^3}
O^{(1)}_{\mu\nu} \mbox{~tr}
\left[\Lambda^{(\pm)}_{p}\gamma^{\mu}
\Lambda^{(\mp)}_{q}\gamma^{\nu}\right] = \pi \Bigg[
\frac{2}{qp}-\frac{(q^2-p^2)^2+\omega_{l}^4}
{\sqrt{3}\omega_{l}^2 q^2 p^2}\arctan
\left(\frac{\sqrt{3}\omega_{l}\mbox{min}(q,p)}
{\omega_{l}^2+|q^2-p^2|-\omega_{l}\mbox{max}(q,p)}\right)
 \nonumber\\
&&-\frac{(q^2-p^2)^2-\omega_{l}^4} {3\omega_{l}^2 q^2 p^2}
\ln\frac{\omega_{l}+|q+p|}{\omega_{l}+|q-p|}
+\frac{(q^2-p^2)^2-\omega_{l}^4} {6\omega_{l}^2 q^2 p^2}
\ln\frac{\omega_{l}^2+|q+p|^2-\omega_{l}|q+p|}
{\omega_{l}^2+|q-p|^2-\omega_{l}|q-p|}+\frac{4}{3qp}
\ln\frac{\omega_{l}^3+|q+p|^3}{\omega_{l}^3+|q-p|^3}
\Bigg],
\label{a-14}
\end{eqnarray}
\begin{eqnarray}
&&\int \frac{ d\Omega}{|\vec{q}-\vec{p}|^2+\omega^2+2M^2}
O^{(2)}_{\mu\nu} \mbox{~tr}
\left[\Lambda^{(\pm)}_{p}\gamma^{\mu}
\Lambda^{(\pm)}_{q}\gamma^{\nu}\right] =\frac{2\pi}{q p}
+\frac{\pi}{2q^2 p^2} \Bigg[
-\frac{(q^2-p^2)^2}{2M^2+\omega^2}
\ln\frac{(q+p)^2}{(q-p)^2}\nonumber\\ &&
-\frac{\left[(q-p)^2+2M^2+\omega^2\right]
\left[(2M^2+\omega^2)^2+(q+p)^2\omega^2\right]}
{2M^2(2M^2+\omega^2)}
\ln\frac{(q+p)^2+2M^2+\omega^2}
{(q-p)^2+2M^2+\omega^2}\nonumber\\ &&
+\frac{[(q+p)^2+\omega^2][(q-p)^2+\omega^2]}{2M^2}
\ln\frac{(q+p)^2+\omega^2}{(q-p)^2+\omega^2} \Bigg],
\label{a-15}
\end{eqnarray}
\begin{eqnarray}
&&\int \frac{ d\Omega}{|\vec{q}-\vec{p}|^2+\omega^2+2M^2}
O^{(2)}_{\mu\nu} \mbox{~tr}
\left[\Lambda^{(\pm)}_{p}\gamma^{\mu}
\Lambda^{(\mp)}_{q}\gamma^{\nu}\right] =-\frac{2\pi}{q p}
+\frac{\pi}{2q^2 p^2} \Bigg[
\frac{(q^2-p^2)^2}{2M^2+\omega^2}
\ln\frac{(q+p)^2}{(q-p)^2}\nonumber\\ &&
+\frac{\left[(q+p)^2+2M^2+\omega^2\right]
\left[(2M^2+\omega^2)^2+(q-p)^2\omega^2\right]}
{2M^2(2M^2+\omega^2)}
\ln\frac{(q+p)^2+2M^2+\omega^2}
{(q-p)^2+2M^2+\omega^2}\nonumber\\
&&-\frac{[(q+p)^2+\omega^2][(q-p)^2+\omega^2]}{2M^2}
\ln\frac{(q+p)^2+\omega^2}{(q-p)^2+\omega^2} \Bigg],
\label{a-16}
\end{eqnarray}
\begin{eqnarray}
 d  \int \frac{ d\Omega }{|\vec{q}-\vec{p}|^2+\omega^2}
O^{(3)}_{\mu\nu} \mbox{~tr}
\left[\Lambda^{(\pm)}_{p}\gamma^{\mu}
\Lambda^{(\pm)}_{q}\gamma^{\nu}\right] &=& \pi  d \Bigg[
-\frac{2}{qp}+\frac{(q+p)^2+\omega^2}{2q^2 p^2}
\ln\frac{(q+p)^2+\omega^2}{(q-p)^2+\omega^2} \Bigg] ,
\label{a-17} \\
 d  \int \frac{ d\Omega }{|\vec{q}-\vec{p}|^2+\omega^2}
O^{(3)}_{\mu\nu} \mbox{~tr}
\left[\Lambda^{(\pm)}_{p}\gamma^{\mu}
\Lambda^{(\mp)}_{q}\gamma^{\nu}\right] &=& \pi  d \Bigg[
\frac{2}{qp}-\frac{(q-p)^2+\omega^2}{2q^2 p^2}
\ln\frac{(q+p)^2+\omega^2}{(q-p)^2+\omega^2} \Bigg].
\label{a-18}
\end{eqnarray}

As a result, the angular average of the gluon propagator
(multiplied by $q^2$ weight) in the vicinity of the Fermi
surface is given by the following approximate expression:
\begin{eqnarray}
q^2 \int d\Omega
{\cal D}_{\mu\nu}(q-p) \mbox{~tr}\left[
\Lambda^{(\pm)}_{p}\gamma^{\mu} 
\Lambda^{(\mp)}_{q}\gamma^{\nu}
\right] &\approx& 2i\pi \left[
\frac{2}{3}\ln\frac{(2\mu)^3}
{|\epsilon_{q}^{-}|^3+\omega_{l}^3}
+\ln\frac{(2\mu)^2}{(\epsilon_{q}^{-})^2+2M^2+\omega^2}
+ d \right] ,
\label{a-19}
\end{eqnarray}
where $\epsilon_{q}^{-}=q-\mu$ and $ d $ if the gauge fixing
parameter. The three terms in the last expression are the
leading contributions of the magnetic, electric and the
longitudinal gluon modes, respectively. Because of the
absence of the logarithmic factor in front of the gauge
fixing parameter, the longitudinal gluon modes become
relevant only in the next-to-leading order.

Now, after performing the approximate integration over the
spatial momentum, we arrive at
\begin{eqnarray}
&&\int_{0}^{\mu}\frac{q^2 dq }
{q_4^2+(\epsilon_{q}^{-})^2+|\Delta|^2}
\int d\Omega  {\cal D}_{\mu\nu}(q-p) \mbox{~tr}\left[
\Lambda^{(\pm)}_{p}\gamma^{\mu} 
\Lambda^{(\mp)}_{q}\gamma^{\nu}
\right] \nonumber\\
&\approx& \frac{4i\pi^2}{3\sqrt{q_4^2+|\Delta|^2}}
\left(\ln\frac{(2\mu)^3}{\omega_{l}^3}
+\frac{3}{2}\ln\frac{(2\mu)^2}{2M^2+\omega^2}
+\frac{3}{2}d\right)
\approx \frac{4i\pi^2}{3\sqrt{q_4^2+|\Delta|^2}}
\ln\frac{\Lambda}{\omega} ,
\label{a-20}
\end{eqnarray}
where $\Lambda\equiv e^{3d/2}(2\mu)^6/(\sqrt{2}\pi M^5)
=e^{3d/2}16 (2\pi)^{3/2}\mu/(N_{f}\alpha_{s})^{5/2}$.

\section{The Solution of the Gap Equation}
\label{appB}

In this Appendix we present a somewhat more rigorous
solution of the integral equation (\ref{gap-q4}).

In order to rewrite it in the form of a differential
equation, we approximate the logarithm in the kernel by its
asymptotes (below $\Delta\equiv \Delta ^{(-)}$,
$p\equiv p_{4}$):

\begin{equation}
\Delta (p)\simeq\frac{2\alpha_{s}}{9\pi}
\int_{0}^{p} \frac{d q \Delta (q)}
{\sqrt{q^2+|\Delta |^2}}
\ln\frac{\Lambda}{p}
+\frac{2\alpha_{s}}{9\pi}
\int_{p}^{\Lambda} \frac{d q \Delta (q)}
{\sqrt{q^2+|\Delta |^2}}
\ln\frac{\Lambda}{q} ,
\label{b-1}
\end{equation}
where, without loss of generality, $\Lambda
=e^{3d/2}(4\pi)^{3/2} \mu/\alpha^{5/2}$, appearing in the
logarithm, is also used as an ultraviolet cut-off in
integration over $q$. This equation, as is easy to check, is
equivalent to the differential equation:
\begin{equation}
p \Delta ^{\prime\prime}(p)+\Delta ^{\prime}(p)
+\frac{2\alpha_{s}}{9\pi}
\frac{\Delta (p)}{\sqrt{p^2+|\Delta |^2}}=0,
\label{b-2}
\end{equation}
along with the following infrared and ultraviolet boundary
conditions:
\begin{equation}
\left. p \Delta ^{\prime}(p)\right|_{p=0}=0
\quad\mbox{(IR)},\qquad
\Delta (\Lambda)=0
\quad\mbox{(UV)}.
\label{b-3}
\end{equation}
We solve the differential equation (\ref{b-2})
analytically  in two regions $p\ll |\Delta |$ and $p\gg
|\Delta |$ and  then match the solutions at
$p=|\Delta |$.

In the infrared region $p\ll |\Delta |$, the solution
that  satisfies the IR boundary condition is given by the
following  Bessel function:
\begin{equation}
\Delta (p)= |\Delta |
J_{0}\left(\nu\sqrt{\frac{p}{|\Delta |}}\right),
\label{b-4}
\end{equation}
where, by definition, $\nu=\sqrt{8\alpha_{s}/9\pi}$ and the
overall constant was chosen in such a way that
$\Delta (0)=|\Delta |$.

In the other region, $p\gg |\Delta |$, the solution,
consistent  with the UV boundary condition, reads
\begin{equation}
\Delta (p)=
B \sin\left(\frac{\nu}{2}\ln\frac{\Lambda}{p}\right).
\label{b-5}
\end{equation}

\begin{figure}
\epsfbox{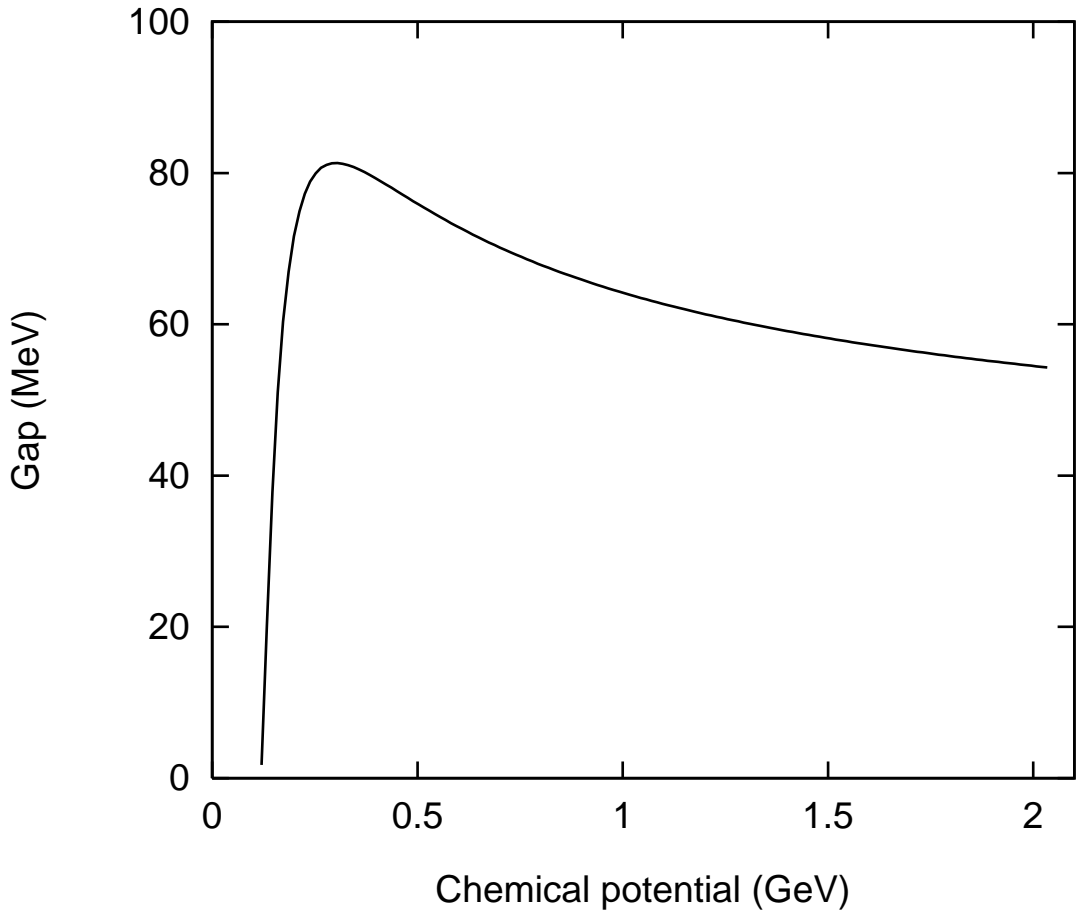}
\caption{The dependence of the order parameter on the
chemical potential. The function $\alpha_{s}(\mu)$ is
established by fixing $\alpha_{s} (1.76 \mbox{GeV})
\approx 0.26$ and using the one-loop running of
the coupling constant.}
\label{fig-del}
\end{figure}

While matching the solutions and their derivatives at the
point $p=|\Delta |$, we get the relation:
\begin{equation}
|\Delta |=\Lambda\exp\left[
-\frac{2}{\nu}\arctan\left(
\frac{J_{0}(\nu)}{J_{1}(\nu)}\right)
\right],
\label{b-6}
\end{equation}
and determine the value of the constant $B$:
\begin{equation}
B=|\Delta |\sqrt{J_{0}^2(\nu)+J_{1}^2(\nu)}.
\label{b-7}
\end{equation}
The dependence (\ref{b-6}) of the order parameter on the
chemical potential is presented in Fig.~\ref{fig-del}.
Here, to establish the function $\alpha_{s}(\mu)$, we fix
the magnitude of the QCD coupling constant by choosing
$\alpha_{s} (1.76 \mbox{GeV})\approx 0.26$. Then, by
making use of the one-loop running of $\alpha_{s}$, we
obtain the functional dependence of $\alpha_{s}$ on
$\mu$.

Finally, assuming that the coupling constant is small,
$\nu\ll 1$, we arrive at the analytical expression presented
in Eq.~(\ref{A-d}):
\begin{equation}
|\Delta |\simeq\Lambda\exp\left(-\frac{\pi}{\nu}
+1+O(\nu^2)\right).
\label{b-8}
\end{equation}


\end{document}